\documentclass[a4paper,11pt]{article}
\pdfoutput=1
\usepackage{jcappub}

\usepackage{tikz,xcolor,hyperref}
\usepackage{amsmath, amssymb, amsthm, graphicx, epsfig, fancyhdr,epsfig, slashed}
\usepackage[normalem]{ulem}
\usepackage{tikzsymbols}
\usepackage{natbib}
\usepackage{float}
\usepackage{placeins}

\def\beq{\begin{equation}}
\def\eeq{\end{equation}}
\def\bea{\begin{eqnarray}}
\def\eea{\end{eqnarray}}
\def\be{\begin{equation}}
\def\ee{\end{equation}}

\def\bse{\begin{subequations}}
\def\ese{\end{subequations}}

\def\Mpl{M_{P}}
\def\f{\frac}
\def\l{\left}
\def\r{\right}

\def\Tbh{T_{\text {BH}}}

\def\tin{t_{\rm in}}

\def\Min{M_{\text {in}}}

\def\mbh{M_{\text {BH}}}

\def\ogw{\Omega_{\text{GW}}}
\def\ast{a_\star}
\def\tin{t_{\rm in}}

\def\ain{a_{\rm in}}

\def\aev{a_\text{ev}}

\def\Tin{T_\text{in}}

\def\gss{g_{\star s}}
\def\gs{g_{\star}}

\def\DNeff{\Delta N_{\rm eff}}

\def\tev{t_{\rm ev}}
\begin{document}
\title{Constraining burdened PBHs with gravitational waves}
\author[a]{Basabendu Barman,}
\emailAdd{basabendu.b@srmap.edu.in}
\author[b]{Kousik Loho}
\emailAdd{kousikloho@hri.res.in}
\author[c]{and Óscar Zapata}
\emailAdd{oalberto.zapata@udea.edu.co}
\affiliation[a]{\,\,Department of Physics, School of Engineering and Sciences, SRM University-AP, Amaravati 522240, India}
\affiliation[b]{\,\,Regional Centre for Accelerator-based Particle Physics,
Harish-Chandra Research Institute, A CI of Homi Bhabha National Institute,
Chhatnag Road, Jhunsi, Prayagraj 211019, India}
\affiliation[c]{\,\,Instituto de Física, Universidad de Antioquia\\Calle 70 \# 52-21, Apartado Aéreo 1226, Medellín, Colombia}
\abstract
{We investigate the implications of memory burden on the gravitational wave (GW) spectrum arising from the Hawking evaporation of light primordial black holes (PBHs). By considering both rotating (Kerr) and non-rotating (Schwarzschild) PBHs, we demonstrate that the overproduction of primordial GWs from burdened PBHs could impose stringent constraints on the parameters governing backreaction effects. These constraints, derived from $\Delta N_{\rm eff}$ measurements by Planck and prospective experiments such as CMB-S4 and CMB-HD, offer novel insights into the impact of memory burden on PBH dynamics.
}
\begin{flushright}
\small{HRI-RECAPP-2024-06}
\end{flushright}
\maketitle
\section{Introduction}
\label{sec:intro}
Since Stephen Hawking's groundbreaking proposals~\cite{Hawking:1974rv, Hawking:1975vcx}, primordial black holes (PBHs) have continuously fascinated astrophysicists and cosmologists. These objects offer a unique opportunity to probe a wide range of intriguing phenomena across several decades in energy, thanks to their extensive mass range, spanning from 0.1 g to $10^5$ solar masses~\cite{Carr:2020gox}. On one hand, PBHs with lifetimes comparable to or longer than the age of the Universe, specifically those with masses $\mbh \gtrsim 10^{15}$ g, could account for the entire dark matter abundance within the mass range $10^{17} \lesssim \Min/\text{g} \lesssim 10^{23}$~\cite{Carr:2020gox, Green:2020jor}. Additionally, these PBHs  may be responsible for the mergers detected through gravitational waves (GWs)~\cite{Bird:2016dcv,DeLuca:2020qqa,DeLuca:2020agl,NANOGrav:2023hvm}, among other possible explanations for different phenomena~\cite{Barack:2018yly,Sasaki:2018dmp}. 
On the other hand, PBHs with masses below $10^9$ g may have played a significant role in the early Universe by dominating the energy density before their evaporation. Moreover,  PBH decay could be crucial in producing SM particles and novel physics states, potentially contributing to dark matter production~\cite{Morrison:2018xla, Gondolo:2020uqv, Bernal:2020bjf, Green:1999yh, Khlopov:2004tn, Dai:2009hx, Allahverdi:2017sks, Lennon:2017tqq, Hooper:2019gtx, Chaudhuri:2020wjo, Masina:2020xhk, Baldes:2020nuv, Bernal:2020ili, Bernal:2020kse, Lacki:2010zf, Boucenna:2017ghj, Adamek:2019gns, Carr:2020mqm, Masina:2021zpu, Bernal:2021bbv, Bernal:2021yyb, Samanta:2021mdm, Sandick:2021gew, Cheek:2021cfe, Cheek:2021odj, Barman:2021ost, Borah:2022iym,Chen:2023lnj,Chen:2023tzd,Kim:2023ixo,Gehrman:2023qjn,Coleppa:2022pnf,Chaudhuri:2023aiv}, baryon asymmetry~\cite{Baumann:2007yr, Hook:2014mla, Fujita:2014hha, Hamada:2016jnq, Morrison:2018xla, Hooper:2020otu, Perez-Gonzalez:2020vnz, Datta:2020bht, JyotiDas:2021shi, Smyth:2021lkn, Barman:2021ost, Bernal:2022pue, Ambrosone:2021lsx,Calabrese:2023key,Calabrese:2023bxz,Gehrman:2022imk,Gehrman:2023esa,Schmitz:2023pfy}, and cogenesis~\cite{Fujita:2014hha, Morrison:2018xla, Hooper:2019gtx, Lunardini:2019zob, Masina:2020xhk, Hooper:2020otu, Datta:2020bht, JyotiDas:2021shi, Schiavone:2021imu, Bernal:2021yyb, Bernal:2021bbv, Bernal:2022swt,Barman:2022pdo,Borah:2024qyo,Chianese:2024nyw}. 

The theoretical framework discussed above, however, rests on the assumption that the black hole remains a classical entity throughout its lifetime, thereby validating the semi-classical approach. It has been demonstrated that Hawking's semi-classical calculations lose their accuracy once the black hole has evaporated approximately half of its original mass~\cite{Dvali:2018xpy,Dvali:2018ytn,Dvali:2020wft}. This is what is known as the {\it memory burden}, which manifests as a universal phenomenon inhibiting further evaporation of the PBH. The phenomenon of memory-burden suggests that the quantum information retained in the memory of a system acts back upon the system (termed as the backreaction), contributing to its stabilization, typically occurring halfway through its decay. In Refs.~\cite{Balaji:2024hpu,Haque:2024eyh,Barman:2024iht} phenomenological implications of memory burden have been explored in context with DM, baryon asymmetry and primordial gravitational wave (GW) due to PBH density perturbation, whereas in Refs.~\cite{Thoss:2024hsr, Dvali:2024hsb} the implications of  PBHs as DM candidates were investigated with memory burden effects taken into account.

The spectrum of GW that are generated by the Hawking evaporation of PBHs has been studied in Refs.~\cite{Anantua:2008am,Dolgov:2011cq,Fujita:2014hha}\footnote{PBHs can also play as the source for primordial GWs in several other ways, e.g., by inducing large curvature perturbation~\cite{Baumann:2007zm,Espinosa:2018eve,Domenech:2019quo,Ragavendra:2020sop,Inomata:2023zup,Franciolini:2023pbf,Firouzjahi:2023lzg,Heydari:2023xts,Heydari:2023rmq,Maity:2024odg}, through merger~\cite{Zagorac:2019ekv} or by the fluctuation of PBH number density~\cite{Domenech:2020ssp,Papanikolaou:2020qtd,Barman:2022pdo,Domenech:2021wkk,Papanikolaou:2022chm,Bhaumik:2022pil,Bhaumik:2022zdd,Balaji:2024hpu}.}. This nearly-thermal GW spectrum is the consequence of graviton emission due to PBH evaporation, and typically peaks at a frequency in the GHz-THz range. Now, non-rotating Schwarzschild black holes predominantly emit particles with lower spin, such as scalars, while the emission of higher-spin particles, like gravitons, is suppressed. In contrast, rotating Kerr PBHs preferentially emit gravitons. This is because the emission of higher-spin modes is amplified, drawing rotational energy from the black hole. This amplification significantly increases the likelihood of graviton emission by several orders of magnitude. The spectrum of primordial GW from a spinning PBH was first computed in Ref.~\cite{Dong:2015yjs}, considering the semi-classical approach to be true. Motivated from these, this study explores the implications of primordial GW emission from both spinning (Kerr) and non-spinning (Schwarzschild) PBHs, incorporating the effects of memory burden\footnote{In Ref.~\cite{Bhaumik:2024qzd} the effect of memory burden on doubly-peaked secondary stochastic gravitational wave background induced by PBH density fluctuation has been discussed, while Ref.~\cite{Kohri:2024qpd} analyzed the memory burden effect on induced GW by primordial curvature
perturbations.}. Given the largely unconstrained nature of memory burden, our findings reveal that the overproduction of primordial GWs from a burdened PBH can impose stringent constraints on the parameters governing the onset of backreaction. We present our results for both PBH and radiation-dominated backgrounds, meticulously accounting for the greybody factors.  

The paper is organized as follows. In Sec.~\ref{sec:framework} we discuss the ingredients for memory burden.  Sec.~\ref{sec:GW} deals with the computation of GW spectrum from PBH evaporation. The outcome of our analysis is presented in Sec.~\ref{sec:result}. Finally, we conclude in Sec.~\ref{sec:concl}.
\section{Beyond the semiclassical regime}
\label{sec:framework}
Let us first briefly discuss the necessary ingredients required to understand the effect of spin and memory burden on PBH dynamics. The mass of the PBH formed due to the gravitational collapse is closely related to the horizon size at the point of formation as~\cite{Fujita:2014hha,Masina:2020xhk}
\begin{align}\label{eq:min}    
\Min=\frac{4}{3}\,\pi\,\gamma\,\left(\frac{1}{H\left(T_\text{in}\right)}\right)^3\,\rho_\text{rad}\left(T_\text{in}\right)\,,
\end{align}
where $H(\Tin)$ corresponds to the Hubble rate during radiation domination at the time of PBH formation. Therefore, radiation temperature at the point of PBH formation
\begin{align}\label{eq:Tin}
& \Tin=\left(\frac{1440\,\gamma^2}{g_\star(\Tin)}\right)^{1/4}\,\left(\frac{M_P}{\Min}\right)^{1/2}\,M_P\,.    
\end{align}
Here, $\gamma$ represents the efficiency factor, which defines what fraction of the total mass inside the Hubble radius collapses to form PBHs. Here $g_\star(\Tin)$ is the relativistic degrees of freedom associated with the thermal bath at the point of formation and $\Mpl= 1/\sqrt{8 \pi G}\simeq 2.4\times 10^{18}$ GeV is the reduced Planck mass. For PBH formation during standard radiation domination $\gamma\simeq 0.2$~\cite{Carr:1974nx}.
The time of the PBH formation is then given by 
\begin{align}\label{eq:tin}
t_{\rm in}=\Min/(8\pi\gamma\Mpl^2)\,,  
\end{align}
where we have considered $H(t)=1/(2t)$ because of standard radiation domination. We parameterize the initial PBH abundance via a dimensionless parameter 
\begin{align}
& \beta\equiv \frac{\rho_{\rm BH}(\tin)}{\rho_R(\tin)}\,,   
\end{align}
which corresponds to the ratio of the initial PBH energy density to the SM energy density at
the time of formation. 

The Kerr black hole is axially symmetric but not spherically symmetric (that is rotationally
symmetric about one axis only, which is the angular-momentum axis), and is characterized by two parameters, namely, the PBH mass $\mbh$ and the dimensionless spin parameter
\begin{align}
a_\star=8\,\pi\,\frac{M_P^2}{\mbh^2}\,J\,, 
\end{align}
where $J$ is the angular momentum and $a_\star\in[0,\,1]$, where $a_\star=0$ corresponds to the non-spinning Schwarzschild black hole.  In Boyer-Lindquist coordinates, the geometry of a Kerr black hole is described by the metric~\cite{Kerr:1963ud}
\begin{align}
ds^2 = -dt^2+\frac{\rho^2}{\Delta}\,dr^2+\frac{2\,r_g\,r}{\rho^2}\,\left(dt-\ast\,r_g\,\sin^2\theta\,d\phi\right)^2+\rho^2\,d\theta^2+\left(r^2+\ast^2\,r_g^2\right)\,\sin^2\theta\,d\phi^2\,,
\end{align}
where
\begin{align}
& \Delta = r^2-2\,r_g\,r+a_\star^2\,r_g^2\,, & \rho^2=r^2+a_\star^2\,r_g^2\,\cos^2\theta\,,    
\end{align}
with $r_g=G\,\mbh$. The coordinates $r,\,\theta,\,\phi,\,t$ are called Boyer–Lindquist coordinates. For a given $\mbh$, corresponding entropy $S$ and horizon  temperature $T_{\rm BH}$, for a spinning PBH is given by 
\begin{align} 
& T_{\rm BH}=\f{2\,\Mpl^2}{\mbh}\,\frac{\sqrt{1-a_\star^2}}{1+\sqrt{1-a_\star^2}}\,,
\nonumber\\&
S(\mbh) =\f{1}{2}\l(\f{\mbh}{\Mpl}\r)^2=\f{1}{2}\l(\f{\Mpl}{T_{\rm BH}}\r)^2\,.
\label{eq:TbhS}
\end{align}
Note that this reduces to the temperature of the 
Schwarzschild black hole for $a_\star=0$, and tends to 0 in the extremal limit $a_\star\to 1$. The Kerr black hole is also characterized by horizon angular velocity, that reads
\begin{align}
\Omega_\star=\frac{4\pi\,a_\star}{1+\sqrt{1-a_\star^2}}\,\frac{M_P^2}{\mbh}\,.    
\end{align}

Once PBH is formed, it can evaporate by emitting Hawking radiation~\cite{Hawking:1974rv, Hawking:1975vcx}. The mass and spin loss rate for PBH can be tracked via following set of equations as~\cite{MacGibbon:1991tj}
\begin{align}\label{eq:dmdtnoMB}
& \f{d\mbh}{dt}=-64\,\pi^2\,\epsilon(\mbh,\,\ast)\,\f{\Mpl^4}{\mbh^2},\\&
\frac{d\ast}{dt}=-64\,\pi^2\,\ast\,\left[\gamma(\mbh,\,\ast)-2\,\epsilon(\mbh,\,\ast)\right]\,\frac{M_P^4}{\mbh^3}\,,\label{eq:dadt}
\end{align}
where, for a given species $j$,
\begin{align}
& \epsilon_j(\mbh,\,\ast)=\frac{g_j}{2\pi^2}\,\int_0^\infty dE\,E\,\sum_{l=s_i}\,\sum_{m=-l}^{l}\frac{d^2 N_{ilm}}{dp\,dt}\,,
\nonumber\\&
\gamma_j(\mbh,\,\ast)=\frac{g_j}{2\pi^2}\,\int_0^\infty dE\,\sum_{l=s_i}\,\sum_{m=-l}^{l}m\,\frac{d^2N_{ilm}}{dp\,dt}\,,
\end{align}
with
\begin{align}\label{eq:d2N}
\frac{d^2N_{ilm}}{dp\,dt}=\frac{\sigma_{s_i}^{lm}(\mbh,\,p,\,\ast)}{e^{\chi_i}-(-1)^{2s_i}}\,\frac{p^3}{E_i(p)}\,.    
\end{align}
Here $l,\,m$ are
the total and axial angular momentum quantum numbers, respectively, and $\chi_i=(E_i-m\,\Omega_\star)/T_{\rm BH}$. The quantity in Eq.~\eqref{eq:d2N} determines the number of $i$ species (with internal degrees of freedom $g_i$ and spin $s_i$) having momenta between $p$ and $p+dp$, emitted by a spinning BH within the interval $t$ and $t+dt$. Here $\sigma_{s_i}^{lm}$ stands for the absorption cross-section (equivalently, the greybody factor), which describes the effects of the centrifugal and gravitational potential on the particle emission~\cite{Hawking:1974rv,Hawking:1975vcx,Page:1976df,Page:1977um}. 

Before the onset of memory burden effects, we can reliably assume that the early stage of Hawking evaporation is well described by purely semiclassical approximations. In that regime, for a non-spinning PBH, the time-evolution of PBH mass can be obtained using Eq.~\eqref{eq:dmdtnoMB},
\begin{align}
& \mbh(t)=\Min[1-\Gamma_{\rm BH}^0(t-t_{\rm in})]^{1/3}\,,    
\end{align}
where 
\begin{align}
\Gamma_{\rm BH}^0=3\,\epsilon\,\Mpl^4/\Min^3 \quad\text{with}\quad\epsilon =\f{27}{4}\f{\pi g_{\star,H}(T_\text{BH})}{480}\,,
\end{align}
is the decay width associated with the PBH evaporation, and $g_{\star,H}(T_\text{BH})$ encodes the degrees of freedom emitted due to PBH evaporation at a temperature $\Tbh$. The  analytical approximation is obtained assuming the geometric optics (GO) limit for the greybody factor $\sigma_{s_i}^{lm}(\mbh,\,p)=27\pi\,G^2\,\mbh^2$. The lifetime $t_{\rm ev}$ of the PBH reads 
\begin{align}
t_{\rm ev}=\f{1}{\Gamma_{\rm BH}^0}
\simeq 2.5\times 10^{-28} \left(\frac{\Min}{1~\rm{g}}\right)^3~\rm{s}\,.
\label{eq:tev1}
\end{align}

To incorporate the effect of memory burden, we consider that the semi-classical regime is valid until 
\begin{align}
\mbh=q\Min\,.    
\end{align}
Since, because of evaporation, the PBH mass diminishes with time, the above relation indicates $q<1$. Beyond the semi-classical regime, the evolution of the PBH mass, as given by Eq.~\eqref{eq:dmdtnoMB}, modifies to~\cite{Dvali:2020wft} 
\bea \label{Eq:memory}
\f{d\mbh}{dt}=-64\,\pi^2\,\f{\epsilon(\mbh,\,\ast)}{\l[S(\mbh)\r]^k}\f{\Mpl^4}{\mbh^2},
\label{eq:dmbhdt}
\eea 
where $S(\mbh)$ is defined in Eq.~\eqref{eq:TbhS}. The exponent $k$ simply decides the efficiency due to the backreaction effect, when the semiclassical approximation breaks down, and we consider it as a free parameter with $k\geq0$.  For a Schwarzschild PBH, this happens at a time~\cite{Haque:2024eyh,Barman:2024iht}
\begin{align}
t_q= \f{1-q^3}{\Gamma_{\rm BH}^0}\,.
\end{align}
Once the mass of the PBH reaches $q\,\Min$, at $t_q$, the quantum memory effect becomes non-negligible. Once the memory effect sets in, total evaporation time for the PBH becomes $\tev=t_q+1/\Gamma^k_{\rm BH}$, where, for a Schwarzschild PBH~\cite{Haque:2024eyh,Barman:2024iht}
\begin{align}\label{eq:Gamma-k}
& \Gamma_{\rm BH}^k= {2^k\,(3+2k)\,\epsilon}\,\Mpl
\l(\f{\Mpl}{q\Min}\r)^{3+2k}\,,    
\end{align}
captures the second phase of evaporation. 
From Eq.~\eqref{eq:Gamma-k}, we find,
\begin{align}
& \tev^k=\frac{1}{\Gamma^k_{\rm BH}}\simeq \frac{1}{2^k\,(3\,\epsilon)}\,\frac{1}{M_P}\,\left(\frac{q\,\Min}{M_P}\right)^3\,, 
\end{align}
for $2k\ll 3$, implying, a larger $k$ results in longer PBH lifetime. Finally, as the radiation energy density evolves as $\rho_R\propto a^{-4}$, while the PBH energy density scales as $a^{-3}$, similar to non-relativistic matter. Therefore, there exist a critical value~\cite{Masina:2020xhk,Haque:2024eyh} 
\begin{align}\label{eq:betac}
&\beta_c=\frac{1}{q}\,\sqrt{\frac{\tin}{\tev}}\,,
\end{align}
above which it is possible that after a time the PBH energy density dominates over the radiation energy density, and a PBH dominated epoch starts until the evaporation completes. For a Schwarzschild PBH of mass 100 g, $\beta_c\simeq 4\times10^{-8}$ for $k=0.1$ and $q=0.5$. 

\begin{figure}[t!]
\centering
\includegraphics[scale=0.5]{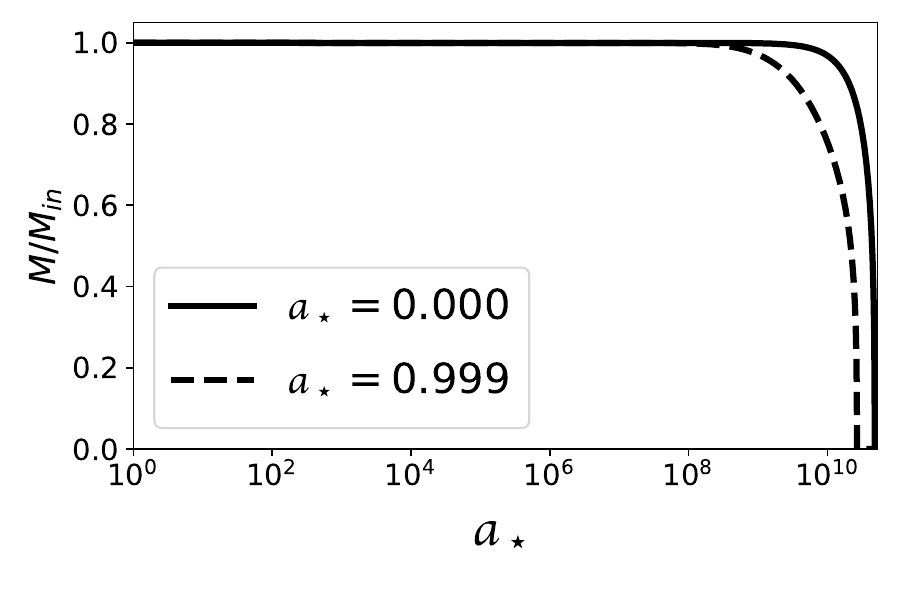}~\includegraphics[scale=0.5]{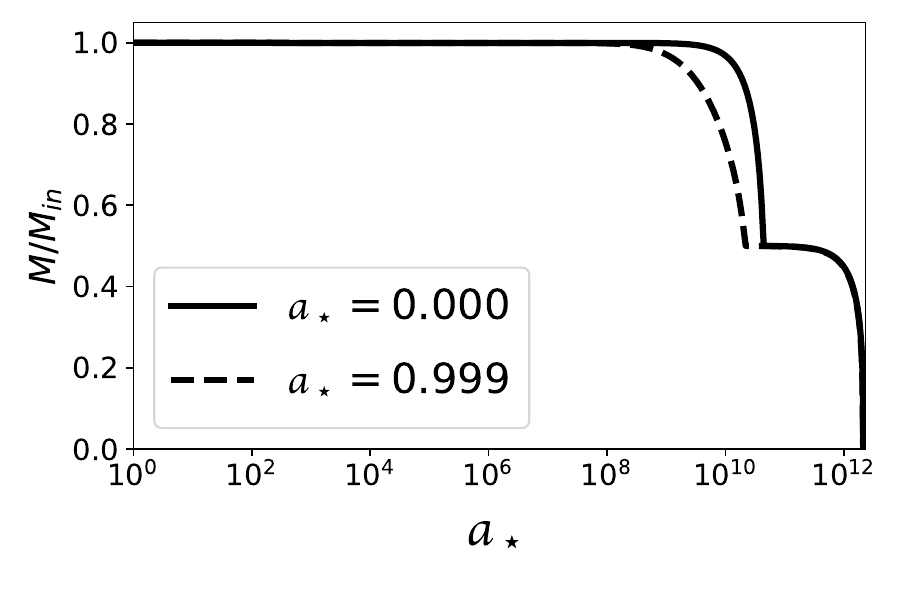}\\[10pt]
\includegraphics[scale=0.5]{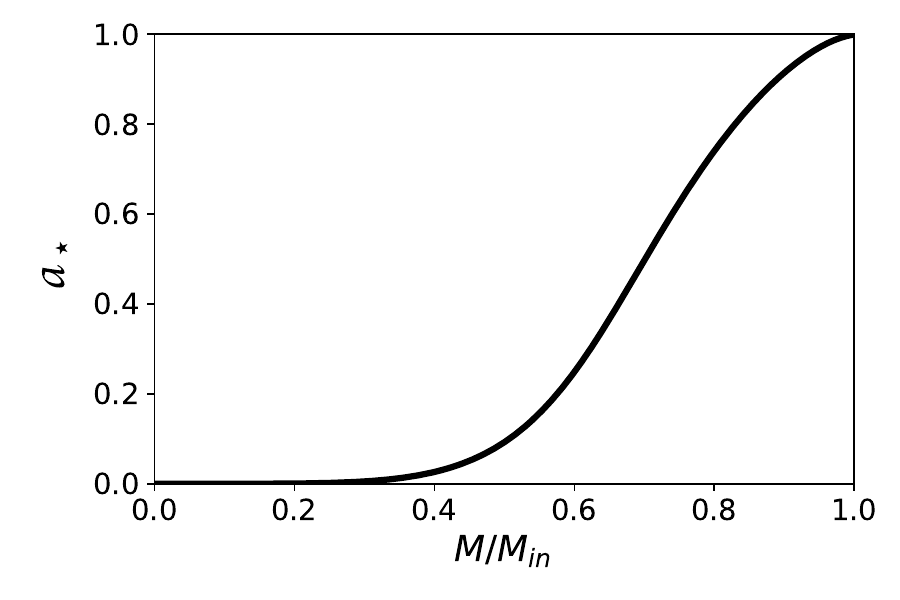}~\includegraphics[scale=0.5]{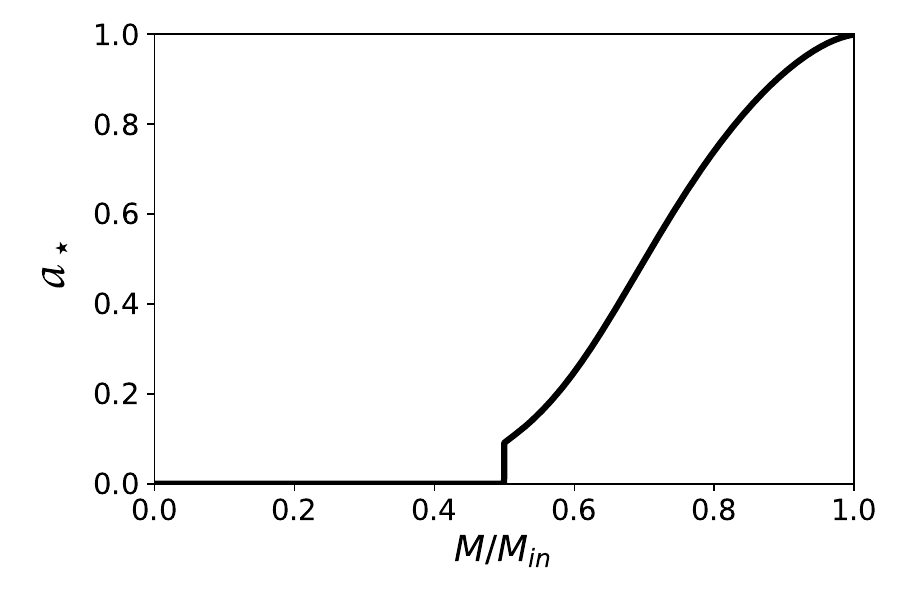}
\caption{Top Left: $\mbh/\Min$ as a function of $a/\ain$ for Schwarzschild (solid line) and nearly-extremal Kerr (dashed line) PBH with $\Min=10^4$ g. Top Right: Same as top left but with memory burden effect included, with  $q=0.5$, $k=0.2$. Bottom Left: Mass and spin loss rate for a Kerr PBH with $\Min=10^4$ g. Bottom Right: Same as bottom left but now including the effect of memory burden with  $q=0.5$, $k=0.2$.}
\label{fig:mass-spin}
\end{figure}

To meticulously track the dynamics of PBHs while accounting for spin and memory burden effects, we have employed the publicly available code {\tt FRISBHEE}~\cite{Cheek:2021odj,Cheek:2021cfe,Cheek:2022dbx,Cheek:2022mmy}. In Fig.~\ref{fig:mass-spin} we present the evolution of PBH mass and spin, with and without the impact of memory burden. We have chosen a benchmark PBH mass of $\Min=10^4$ g, however we stress that the observed characteristics are applicable for any PBH mass. Furthermore, we assume that the PBHs possess a {\it monochromatic} mass (and spin) distribution. The top left panel shows the effect of spin on PBH mass. As one can see, both Schwarzschild and Kerr PBH evolve identically for $\mbh/\Min\lesssim 0.4$, typically when the Kerr PBH has shaded most of its angular momentum and starts evolving like a Schwarzschild, as one can see by comparing with the bottom left panel. Following this, the PBH evaporates almost instantaneously, with its mass rapidly dropping to zero. The inclusion of spin hastens the evaporation process, as indicated by the dashed curve, resulting in a shorter lifetime for Kerr PBHs. When the memory burden is considered, a second phase of evolution becomes clearly visible at $\mbh=q\,\Min$, regardless of spin, as shown in the top right and in the bottom right panel. However, the total evaporation time for the PBHs is identical, because once the Kerr PBH spins down, it evolves same as a Schwarzschild PBH, with lifetime simply following Eq.~\eqref{eq:Gamma-k}.

Before closing this section let us comment on the allowed mass range for PBH. The minimum mass is set by the scale of inflation, which for Schwarzschild PBH is roughly around $\sim 0.5$ g, calculated assuming de-Sitter-like inflation considering the tensor to scale ratio $r<0.036$ from Planck~\cite{Planck:2018jri}, together with the latest BICEP/$Keck$ data~\cite{Planck:2018jri,BICEP:2021xfz,Tristram:2021tvh}. This implies $k\lesssim3$ for complete evaporation before BBN.
\section{Gravitational wave from PBH evaporation}
\label{sec:GW}
Once PBHs complete their evaporation, the particles they emit are typically thermalized. However, gravitons are a notable exception. Due to their extremely suppressed interactions, gravitons retain their original momentum distribution, meaning the spectrum of gravitational waves (GW) preserves crucial information about the PBHs even after they have evaporated. This preserved spectrum serves as a unique signature, potentially offering insights into the properties and existence of the PBHs. The GW signal for a monochromatic mass spectrum with $\Min\sim 1\,(10^4)$ g is expected to peak at frequencies of order $10^{13} \,(10^{15})$ Hz. While there are currently no facilities that are capable of detecting such high-frequency GW, however there are future prospects that can potentially be sensitive to GHz-THz spectrum~\cite{Aggarwal:2020olq,Ringwald:2020ist,Ringwald:2022xif}. A generic characteristic of GW due to graviton production from PBH evaporation is that it is nearly a blackbody spectrum, but features more power in the lower frequency domain.

The fractional energy density in stochastic GW generated by the evaporation of PBHs at present epoch reads (see Appendix.~\ref{sec:app-gw} for derivation in case of Schwarzschild black hole)~\cite{Dong:2015yjs,Ireland:2023avg}
\begin{align}\label{eq:ogw}
& \ogw(\omega_0)=\frac{\omega_0^2\,a_{\rm in}^3}{6\,\pi\,M_P^2\,H_0^2}\,n_{\rm BH}^{\rm in}
\int_{a_{\rm in}}^{a_{\rm ev}}\,\frac{da}{a^2\,H}\,\mathcal{Q}(\ast(t),\,\omega\,\mbh(t))\,, 
\end{align}
where $\omega_0=2\pi\,f_0$ stands for GW frequency and $H_0\simeq 100\, h$ km/s/Mpc is the Hubble rate at the present epoch. The instantaneous graviton spectrum is related to $\mathcal{Q}$-factor as
\begin{align}
&\frac{dE}{dt\,d\omega}=\frac{\omega}{2\pi}\,\mathcal{Q}(\ast(t),\,\omega\,\mbh(t))\,,    
\end{align}
and the initial PBH number density, 
\begin{align}\label{eq:nbhin}
& n_{\rm BH}^{\rm in}=48\,\beta\,\,\pi^2\,\gamma^2\frac{M_P^6}{\Min^3}\,,    
\end{align}
is obtained from Eq.~\eqref{eq:min} and \eqref{eq:Tin}. The lower and upper limit of the integration in Eq.~\eqref{eq:ogw} corresponds to the scale factor at the time of PBH formation
\begin{align}\label{eq:ain}
& \ain=\aev\,\sqrt{\frac{t_{\rm in}}{t_{\rm eq}}}\,\left(\frac{t_{\rm eq}}{t_{\rm ev}}\right)^{2/3}    
\end{align}
and that at the time of PBH evaporation 
\begin{align}\label{eq:aev}
& \aev=a_{\rm BBN}\,\left(\frac{\gs(T_{\rm BBN})}{\gs(T_{\rm ev})}\right)^{1/3}\,\frac{T_{\rm BBN}}{T_{\rm ev}}\,,    
\end{align}
respectively. Here,
\begin{align}\label{eq:teq}
& t_{\rm eq}=\beta^{-2}\,t_{\rm in}\,, 
\end{align}
corresponds to the time when the PBH and radiation energy density becomes equal to each other, resulting the onset of a matter-dominated Universe for $t_{\rm eq}<t<t_{\rm ev}$, during which $a(t)\propto t^{2/3}$. Prior to that, i.e., for $t_i<t<t_{\rm eq}$, the Universe is radiation dominated with $a(t)\propto t^{1/2}$. Further, we assume that the PBHs evaporate before the onset of BBN and $a_{\rm BBN}=a_0\,\left(\gss(T_0)/\gss(T_{\rm BBN})\right)^{1/3}\,\left(T_0/T_{\rm BBN}\right)$, with $T_{\rm BBN}\simeq 4$ MeV. 

Away from the source, the energy density of GW background scales same as that of free radiation energy density, i.e., $a^{-4}$. This implies that the GWs act as an additional source of radiation. Now, an upper bound on any extra radiation component, in addition to those of the SM, can be expressed in terms of the $\DNeff$, defined as
\begin{equation}
    \rho_\text{rad}(T\ll m_e) = \rho_\gamma + \rho_\nu + \rho_\text{GW} = \left[1 + \frac78 \left(\frac{T_\nu}{T_\gamma}\right)^4 N_\text{eff}\right] \rho_\gamma\,,
\end{equation}
where $\rho_\gamma$, $\rho_\nu$, and $\rho_\text{GW}$ correspond to the photon, SM neutrino, and GW energy densities, respectively, with $T_\nu/T_\gamma = (4/11)^{1/3}$. Within the SM, taking the non-instantaneous neutrino decoupling into account, $N_\text{eff}^\text{SM} = 3.044$~\cite{Dodelson:1992km, Hannestad:1995rs, Dolgov:1997mb, Mangano:2005cc, deSalas:2016ztq, EscuderoAbenza:2020cmq, Akita:2020szl, Froustey:2020mcq, Bennett:2020zkv}, while the presence of extra (dark) radiation component in terms of GW can be accommodated in terms of $\DNeff$, defined as 
\begin{align}
& \DNeff = N_\text{eff}-N_\text{eff}^\text{SM} = \frac{8}{7}\,\left(\frac{11}{4}\right)^\frac{4}{3}\,\left(\frac{\rho_\text{GW}(T)}{\rho_\gamma(T)}\right)\,.
\end{align}
Since $\rho_\text{GW}\propto a^{-4}$ and $\rho_\gamma\propto T^4$, the above relation can be utilized to put a constraint on the GW energy density $\rho_{\rm GW}$ redshifted to today via
\begin{align}\label{eq:GW-neff}
& \left(\frac{\rho_\text{GW}}{\rho_c}\right)\Bigg|_0\leq\Omega_{\gamma,0}\,\left(\frac{4}{11}\right)^\frac{4}{3}\,\frac{7}{8}\,\DNeff \simeq 5.62\times 10^{-6}\,\DNeff\,,
\end{align}
where, $\Omega_{\gamma,0}\,h^2\simeq 2.47\times 10^{-5}$ is the fractional photon energy density at present. The bound in Eq.~\eqref{eq:GW-neff} is applicable for an integrated energy density as,
\begin{align}\label{eq:GW-neff2}
&  \left(\frac{h^2\,\rho_\text{GW}}{\rho_c}\right)\Bigg|_0 = \int\,\frac{df_0}{f_0}\,h^2\,\ogw(f_0)\lesssim 5.62\times 10^{-6}\,\DNeff\,, 
\end{align}
where the integration runs from the epoch of PBH formation $\tin$ to evaporation $\tev$. Except for GW spectrum with a very narrow peak of width $\Delta f\ll f_0$, the bound in Eq.~\eqref{eq:GW-neff2} can be interpreted as a bound on the amplitude of GW spectrum itself,
\begin{align}\label{eq:GW-neff3}
& \ogw\,h^2(f_0)\lesssim 5.62\times 10^{-6}\,\DNeff\,,   
\end{align}
over all frequencies. As it is understandable, since the GW spectral energy density depends on the parameters $q,\,k$  that control the effects of memory burden through Eq.~\eqref{eq:ogw}, hence Eq.~\eqref{eq:GW-neff3} can be utilized to obtain bounds on $q$ and $k$. 

Before moving on, let us briefly summarize existing and future bounds on $\DNeff$ from several different experiments. In the context of the $\Lambda$CDM model, the Planck legacy data yields $N_\text{eff} = 2.99 \pm 0.34$ at a 95\% confidence level~\cite{Planck:2018vyg}. When baryon acoustic oscillation (BAO) data are incorporated, this estimate becomes more precise, refining to $N_\text{eff} = 2.99 \pm 0.17$. As detailed in Ref.~\cite{Yeh:2022heq}, a combined analysis of BBN and CMB data reveals $N_\text{eff} = 2.880 \pm 0.144$. Future CMB experiments, such as CMB-S4~\cite{Abazajian:2019eic}, CMB-HD~\cite{CMB-HD:2022bsz} and CMB-Bh$\bar{a}$rat~\cite{Adak:2021lbu}, are expected to achieve sensitivities around $\DNeff \simeq 0.06$, $\DNeff \simeq 0.027$ and $\DNeff\simeq 0.05$, respectively. Additionally, the next generation of satellite missions, like COrE~\cite{COrE:2011bfs} and Euclid~\cite{EUCLID:2011zbd}, promises to further tighten these constraints, potentially reaching a precision of $\DNeff \lesssim 0.013$. In the next section we will utilize these bounds in order to constrain the memory burden parameters by exploiting Eq.~\eqref{eq:GW-neff3}.  
\section{Results and discussions}
\label{sec:result}
\begin{figure}[htb!]
\centering
\includegraphics[scale=0.5]{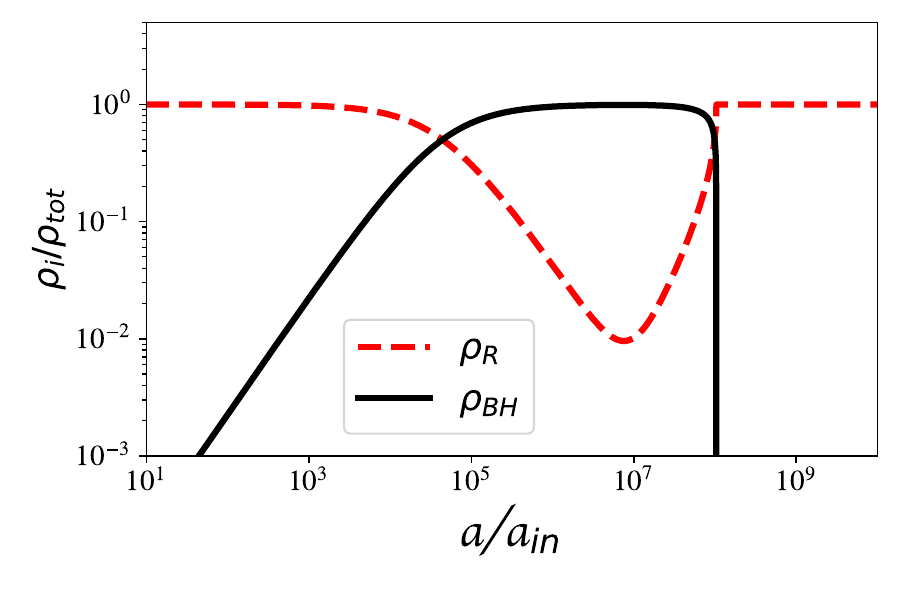}
\includegraphics[scale=0.5]{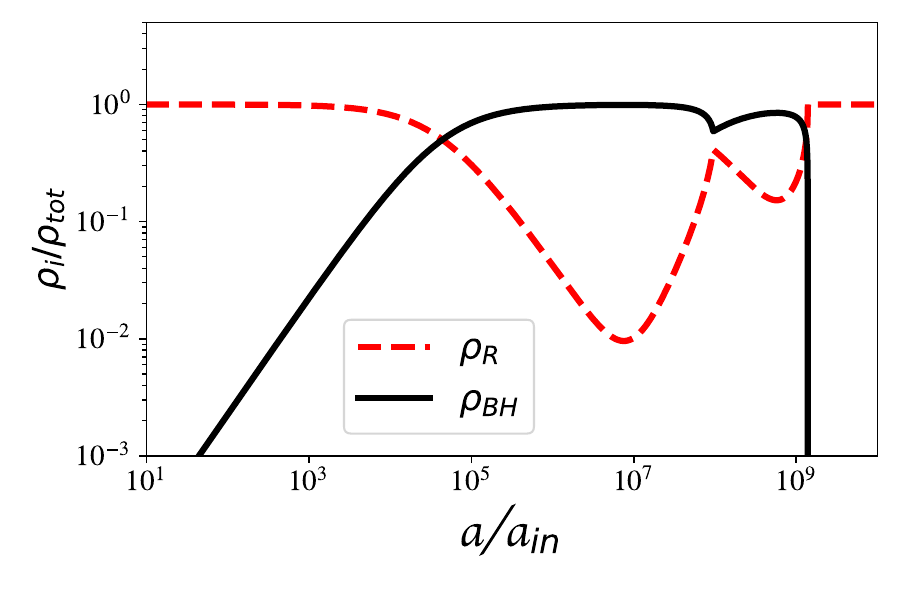}\\
\includegraphics[scale=0.5]{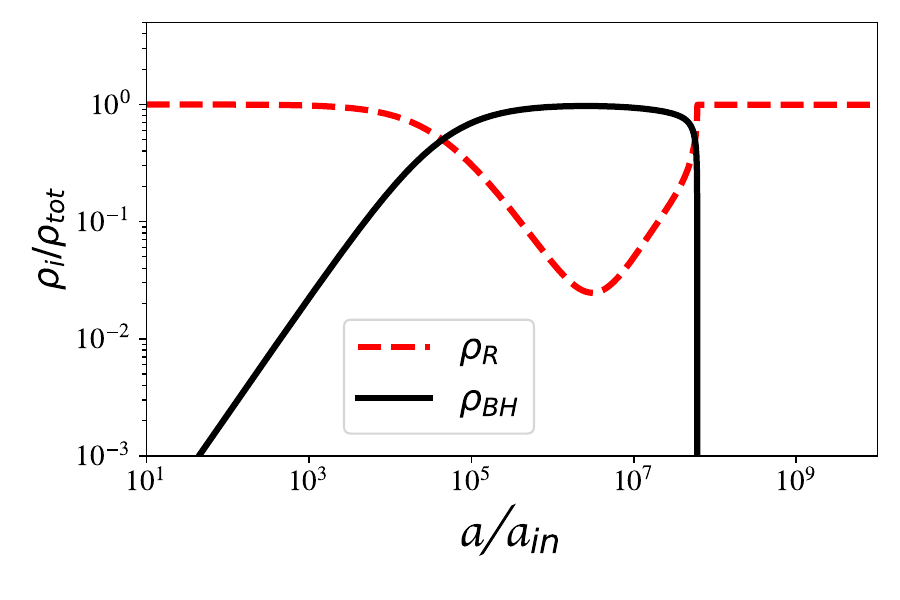}
\includegraphics[scale=0.5]{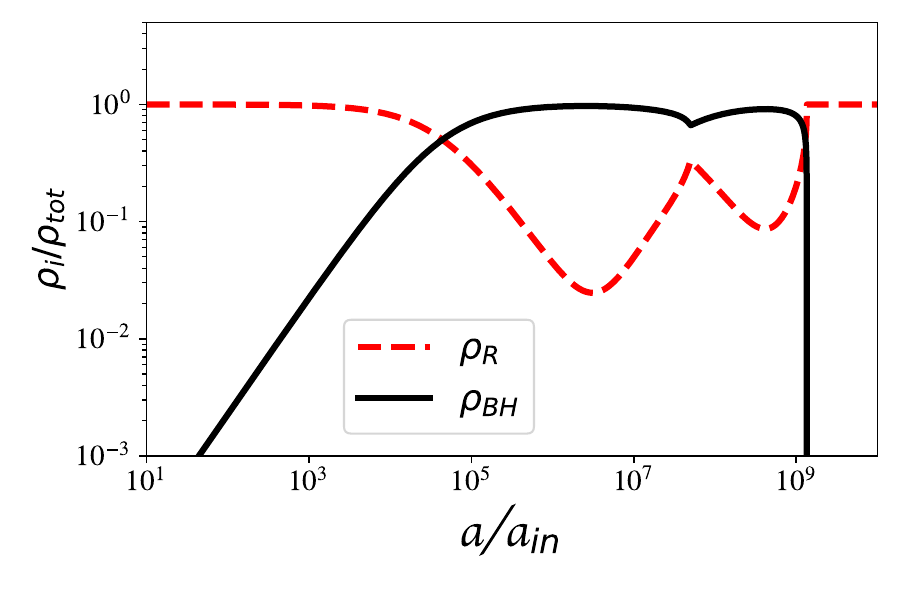}
\caption{Top: Evolution of PBH (black) and radiation (red, dashed) energy densities (scaled to total energy density), as a function of $a/\ain$ for a Schwarzschild PBH, where in the {\it left panel} we do not consider quantum effects due to memory burden, while the for the {\it right panel} we choose $k=0.2,\,q=0.5$ as a benchmark value.
Bottom: Same as top, but now considering Kerr PBH with $\ast=0.99$. In all cases we choose $M_{\rm in}=10^2$ g, $\beta=10^{-5}$.}
\label{fig:evol}
\end{figure}

We begin our discussion by demonstrating the impact of memory burden on the evolution of  PBH and radiation energy densities in Fig.~\ref{fig:evol}, considering Schwarzschild and Kerr PBH. The time evolution of the PBH $(\rho_{\rm BH})$ and SM radiation $(\rho_R)$ energy densities is governed by the Friedmann equations  
\begin{eqnarray}
    \frac{d\rho_{\rm BH}}{dt}+3H\rho_{\rm BH}&=&+\frac{\rho_{\rm BH}}{\mbh}\,\frac{d\mbh}{dt}\,,\\
    \frac{d\rho_R}{dt}+4H\rho_R&=&-\frac{\rho_{\rm BH}}{\mbh}\,\frac{d\mbh}{dt}\,,
\end{eqnarray}
where $H^2=(\rho_{\rm BH}+\rho_R)/(3M_P^2)$, with $d\mbh/dt$ given in Eq.~(\ref{Eq:memory}).   
Here we take a benchmark PBH mass $\Min=100$~g and choose $\beta=10^{-5}$ such that the PBH evaporates during PBH domination [cf.Eq.~\eqref{eq:betac}]. The top left panel shows the scenario where the semi-classical approximation is assumed to be always valid for a non-spinning PBH. Here we see brief period of PBH domination, which eventually ends with the PBH evaporation. The effect of memory burden for the same PBH is illustrated in the top right panel for $q=0.5,\,k=0.2$. There are two distinguishable features: (i) the presence of a second phase of PBH evolution\footnote{For $q>0.41$, PBH domination continues until the time of evaporation~\cite{Balaji:2024hpu}.} due to the breakdown of semi-classical treatment at $\mbh=q\,\Min$, and (ii) substantially longer PBH lifetime. The later is important for GW spectrum as we shall discuss in a moment. In the bottom panel on top of the memory burden effect, we also include the PBH spin. Although this is not very apparent from the bottom panel because of the logarithmic scales, but the spin of PBH shortens its lifespan, with rapidly rotating black holes evaporating more swiftly than their non-rotating counterparts.
\begin{figure}[tb!]
\centering
\includegraphics[scale=0.5]{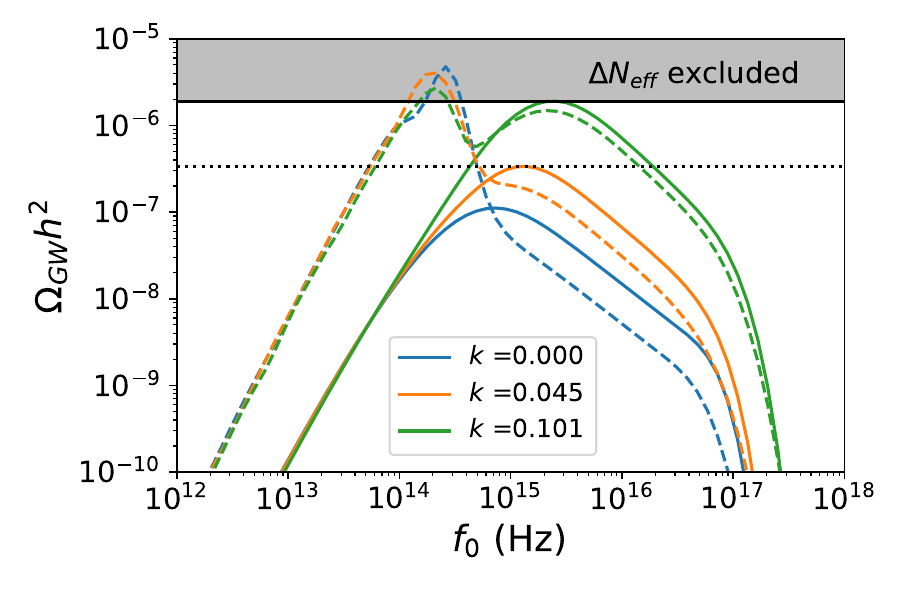}
\includegraphics[scale=0.5]{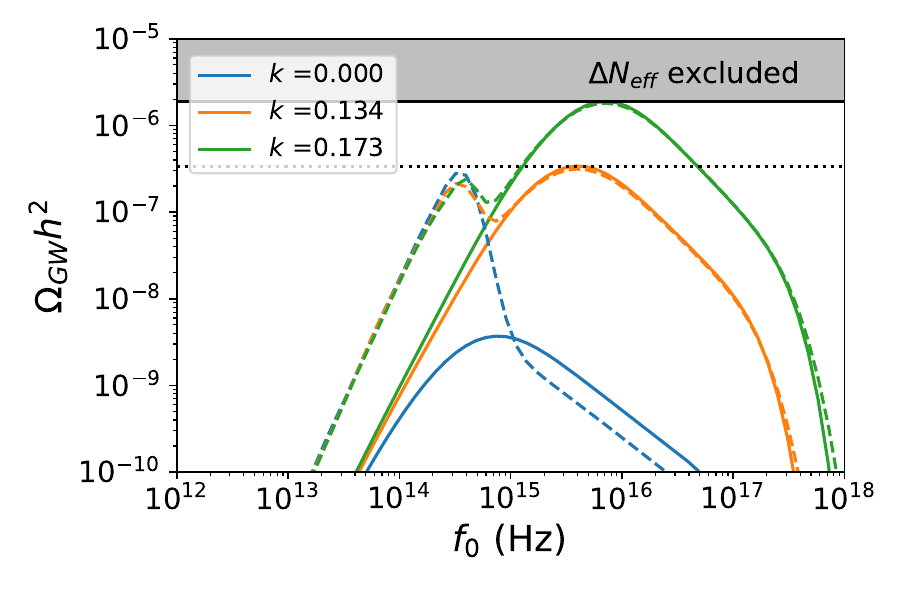}
\caption{Left: Spectrum of GW energy density, as a function of frequency $f_0$ for
$\Min=10^2$ g, $\beta=10^{-5}$ and $q=0.5$ . Colored solid (dashed) lines correspond to Schwarzschild (Kerr) PBH (with $a_\star=0.99$). We choose different $k$-values to demonstrate the effect of memory burden. Right: Same as left for $\beta=10^{-9}$. The black solid and dotted horizontal lines correspond to $\DNeff$ bound due to Planck~\cite{Planck:2018jri} and the sensitivity limit of CMB-S4~\cite{Abazajian:2016hbv,Abazajian:2019eic}, respectively. }
\label{fig:GW}
\end{figure}
\begin{figure}[htb!]
\centering
\includegraphics[scale=0.5]{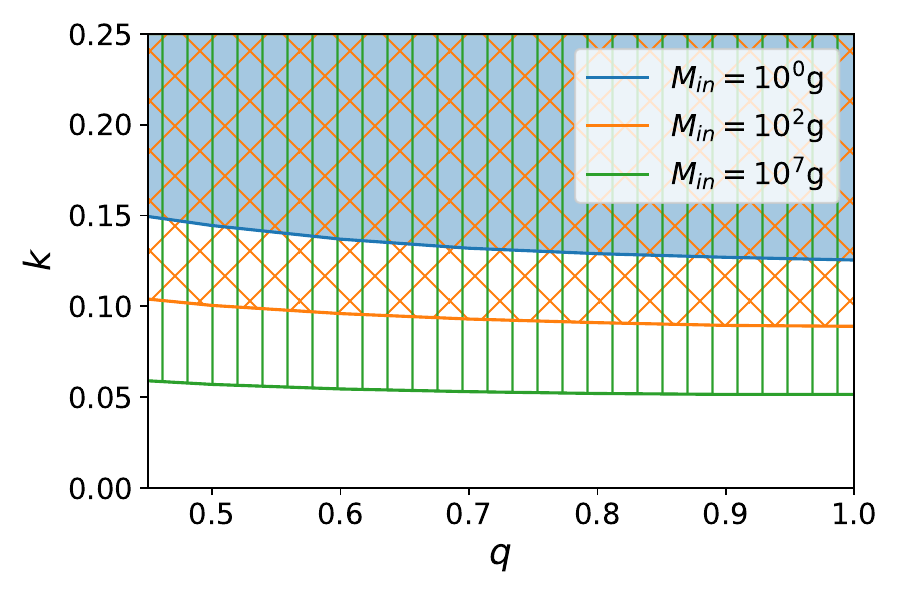}~\includegraphics[scale=0.5]{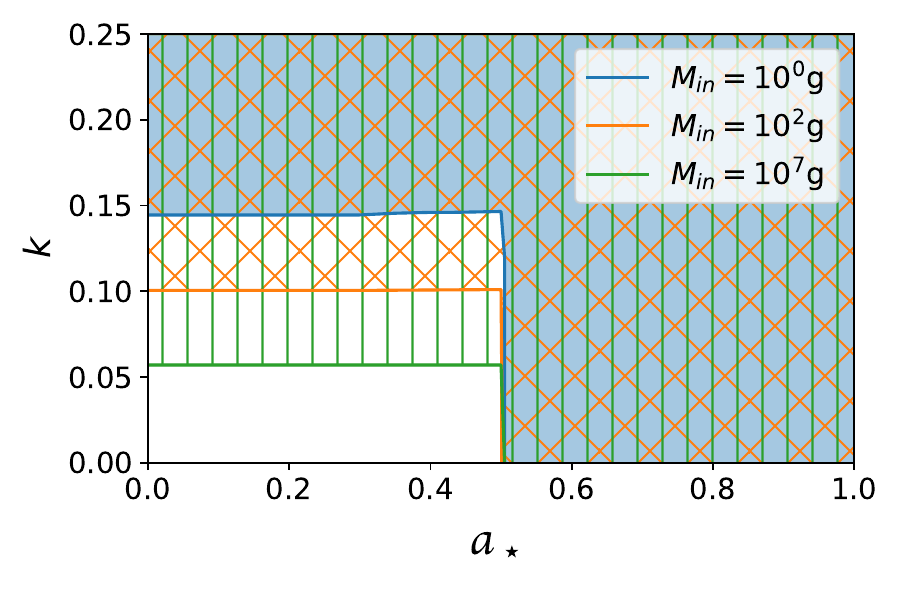}
\caption{Left: Constraints in $[k-q]$ parameter space due to $\DNeff$ from Planck~\cite{Planck:2018vyg} measurement ($\DNeff=0.34$), for Schwarzschild PBH with different choices of the initial mass: $\Min= \{10^0,\,10^2,\,10^7\}$ g. Right: Constraint on $k$ from $\DNeff$ for a spinning PBH for the same set of PBH masses as in the left panel. In either cases we choose $\beta=10^{-4}$ to ensure PBH domination. The shaded region is excluded from overproductionn of GW.}
\label{fig:qvsk}
\end{figure}

The GW spectral energy density $\ogw$, as a function of the linear frequency $f$, is shown in Fig.~\ref{fig:GW} for Schwarzschild (solid lines) and Kerr (dashed lines) PBH. In both cases we include the effect of memory burden and compare with corresponding spectrum obtained assuming the semi-classical estimation to be true. Remarkably, even for $k\ll 1$, the peak amplitude lies within the sensitivity reach of Planck measurement. The current upper bound $\DNeff\simeq 0.34$~\cite{Planck:2018jri} thus  constrains $k\lesssim0.1$ for a Schwarzschild PBH of mass 100 g. Spinning PBHs have a reduced lifetime compared to non-spinning PBHs. As a consequence, the GW spectrum from the spinning PBH experiences a longer period of cosmological redshift. Hence, for near-extremal PBHs, the peak position shifts to lower frequencies, while the amplitude of the peak is enhanced, as PBH spin aids to the emission of higher spin species. Due to this enhancement, for Kerr PBH, even the $k=0$ case is in conflict with existing bound on $\DNeff$ from Planck for $\beta=10^{-5}$ and $\Min=100$ g. This bound can be significantly relaxed for smaller $\beta$, that results in PBH evaporation during radiation domination, as shown in the right panel. One can further notice that for higher values of $k$ and the subsequent delayed evaporation the peaks corresponding to the redshift are also suppressed for Kerr black holes. We note, for Schwarzschild PBH with $k=0$, the spectral energy density peaks at $\omega_0\simeq 2.8\,\aev\,\Tbh$, that is 
\begin{align}\label{eq:fpeak}
    f_0^{\rm peak}\sim2\times10^{16}\,{\rm Hz}\left(\frac{\Min}{10^5\,{\rm g}} \right)^{1/2},
\end{align}
leading to $\ogw^{\rm peak} h^2\sim 10^{-7}$, which is beyond the reach of present observation from Planck as well as the future sensitivity projection from CMB-S4 on $\DNeff$. It is important to note that for a given PBH mass, irrespective of spin, memory effects result in excessive production of GW energy density, resulting in tighter constraint from $\DNeff$ measurements. This is expected because the PBH lifetime always increases with $k$ [cf. Eq.~\eqref{eq:Gamma-k} for Schwarzschild PBH], thereby delaying its evaporation. As a consequence, the gravitons are produced for a longer time, contributing to the GW energy density. 
\begin{figure}[htb!]
\centering
\includegraphics[scale=0.7]{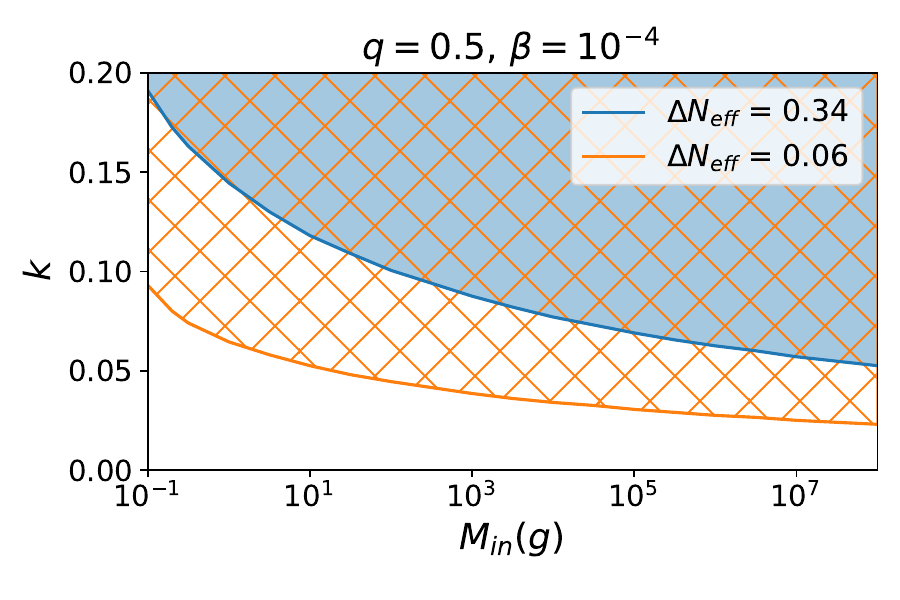}
\caption{Constraints on $k$ as a function of $\Min$ for the benchmark choice $q=0.5$ and $\beta=10^{-4}$. The blue shaded region is excluded from corresponding $\DNeff$ due to Planck~\cite{Planck:2018vyg} whereas the shaded region in orange will be explored by CMB-S4~\cite{Abazajian:2019eic} (in orange).}
\label{fig:Mvsk}
\end{figure}
\begin{figure}[htb!]
\centering
\includegraphics[scale=0.7]{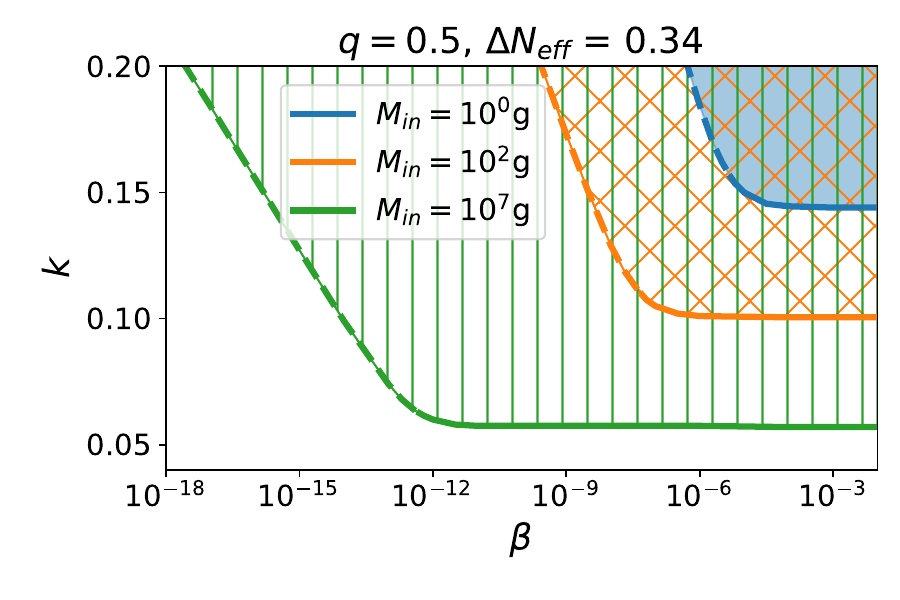}
\caption{Constraints from $\DNeff$ due to Planck~\cite{Planck:2018vyg} on $k$, as a function of $\beta$ for $q=0.5$ and for $M_{in}=\{10^0,\,10^2,\,10^7\}$ g. The shaded region is excluded from overproduction of GW. The absence of PBH domination at smaller values of $\beta$ is represented by dashed lines.}
\label{fig:betavsk}
\end{figure}

Since heavier PBHs have a larger lifetime, hence the constraint on $k$ from $\DNeff$ becomes more stringent because they produce more gravitons over time. This is clearly seen from both panels in Fig.~\ref{fig:qvsk}. However, for a given PBH mass, the bound is almost insensitive to $q$, as seen from the left panel. As we have already noticed in Fig.~\ref{fig:GW}, near-extremal spinning PBHs are already in tension with $\DNeff$ bound. This is once again reflected here, typically more pronounced for higher masses and $\ast\gtrsim 0.5$. As an instance, present bound on $\DNeff$ from Planck allows $k\lesssim 0.05$ and $\ast\lesssim 0.5$ for a PBH of mass $\Min=10^7$ g, with $\beta=10^{-4}$. For comparatively lighter PBHs, the bound on $k$ is relaxed, but near-extremal PBHs are ruled out. 

The effect of PBH mass is more prominent from  Fig.~\ref{fig:Mvsk}, where we vary PBH mass $\Min$ over a large range for a given $q$, considering PBH domination. Again we see, for heavier PBHs, the bound on $k$ from $\DNeff$ is stronger due to, as stated before, longer lifetime that results in more graviton production, enhancing the GW amplitude. This bound becomes even stronger for Kerr PBH, and typically for near-extremal case as we have already seen in Fig.~\ref{fig:GW} and \ref{fig:qvsk}, hence we refrain from showing them here again. Clearly, projected sensitivity reach of CMB-S4 (in orange) improves the bound on $k$ by at least an order for heavier PBHs.

To this end we have been concentrating only on PBH-dominated scenario, where we chose $\beta>\beta_c$. Now, from Eq.~\eqref{eq:Gamma-k} and \eqref{eq:teq}, for a Schwarzschild PBH, we find, PBHs dominate before the evaporation epoch if
\begin{align}\label{eq:betac2}
& \beta>\frac{1}{2}\,\sqrt{\frac{2^{k-1}\,\epsilon}{\pi\,\gamma}\,\frac{2k+3}{q}}\,\left(\frac{M_P}{q\,\Min}\right)^{k+1}\simeq 2.4\times 10^{-6}\,\left(\frac{1\,\text{g}}{\Min}\right)\,,    
\end{align}
for $q=0.5,\,k=0.2$, else, $\ogw$ does not depend on $\beta$. As a result, we see from Fig.~\ref{fig:betavsk}, for a PBH of mass 1 g, the blue solid line becomes parallel to the horizontal axis for $\beta\gtrsim 10^{-6}$. For PBHs with mass $\Min>1$ g, this characteristic remains unchanged, however, PBH domination begins from comparatively smaller $\beta$, as evident from Eq.~\eqref{eq:betac2}. For heavier PBHs\footnote{Overproduction of induced GWs by the density fluctuations due to inhomogeneous distribution of PBHs demands $\beta\lesssim 10^{-3}$, which becomes stronger as the PBH mass increases~\cite{Papanikolaou:2020qtd,Domenech:2020ssp}.}, the bound on $k$ is stronger as we have already realized in Fig.~\ref{fig:Mvsk}. For $t_{\rm eq}<1/\Gamma^k_{\rm BH}$, the PBHs evaporate during radiation domination. As one can see, the constraint on $k$ is comparatively loosened in this case since $\ogw\propto\beta$ [cf.Eq.~\eqref{eq:ogw}]. This is what we already realized in the right panel of Fig.~\ref{fig:GW}. However, as longer $\tev$ (because of memory burden) results in larger graviton flux, it is still not possible to have $k\gtrsim 0.2$. Here we consider a Schwarzschild PBH, while the bound on Kerr PBH, as we have already seen from Fig.~\ref{fig:GW} and \ref{fig:Mvsk}, is very strong, unless $\ast\lesssim 0.5$. Once again, future measurements from, for example, CMB-S4, can improve this bound further, as already noticed in Fig.~\ref{fig:Mvsk}.
\section{Conclusion}
\label{sec:concl}
When the mass of a black hole (BH) reaches a certain fraction $q$ of its initial value, the backreaction can no longer be ignored, and it can potentially reduce the evaporation rate by the inverse power law of its entropy $S^{-k}$. From this point onward the system is affected by what is called the memory burden effect that significantly
enhances the BH lifetime. There is apparently, no as such theoretical constraint on $q$ and $k$, apart from the fact that $k>0$ and $q<1$. In this work we have shown that the primordial gravitational wave (GW), produced entirely from the evaporation of a burdened primordial black hole (PBH), can provide stringent bound on these parameters. This is attributed to the improved lifetime for PBHs beyond semi-classical regime, that results in an enhanced GW amplitude [cf. Fig.~\ref{fig:GW}] which overproduces relativistic degrees of freedom around the time of big bang nucleosynthesis (BBN), thus conflicting the $\DNeff$ constraint. We explore both the case of non-spinning and spinning PBHs, where for the later scenario the probability of getting higher spin particles (for example, graviton) from PBH evaporation is also higher. We find, the bounds are very sensitive to typically the choice of $k$ and $\beta$ and stronger for heavier PBHs as they live longer [cf.~Fig.~\ref{fig:Mvsk} and \ref{fig:betavsk}]. Our study shows, the bounds on the memory burden parameters are stronger for PBH evaporation during PBH domination, but they become comparatively weaker (but non-negligible) for PBH evaporation during radiation domination. Notably, Planck constraint on $\DNeff$ has excluded any potential memory burden effect in case of Kerr PBHs with $\ast>0.5$, for PBH domination [cf. right panel of Fig.~\ref{fig:qvsk}]. Future CMB observations from experiments, for instance, CMB-S4 or CMB-HD, shall further improve this bound, making the period of memory burden more precise.
\section*{Acknowledgment}
B.B. would like to acknowledge fruitful discussions with Md Riajul Haque. B.B. and K.L. would like to acknowledge the hospitality of Institute of Physics (IOP), Bhubaneswar during the ``Workshop on Dark Matter and Astroparticle Physics (WDMAP)", where a part of the project was completed. O.Z. has been partially supported by Sostenibilidad-UdeA, the UdeA/CODI Grants 2022-52380 and  2023-59130, and Ministerio de Ciencias Grant CD 82315 CT ICETEX 2021-1080. 
\appendix
\section{Gravitational wave spectrum from a Schwarzschild PBH}
\label{sec:app-gw}
The differential energy radiated due to PBH evaporation per unit time per unit frequency is given by, 
\begin{align}
&\frac{dE}{dt\,df}=\frac{2\,\pi^2\,g\,f^3}{\exp(2\pi\,f/\Tbh)-1}\,,    
\end{align}
where $\Tbh=M_P^2/\mbh$ is the BH temperature and $g$ is the degree of freedom of the emitted particle species. Now, at present epoch $t=t_0$, the energy density of graviton emitted between frequency $f$ and $f+df$,
\begin{align}
& \frac{g_h}{g}\times n_{\rm PBH}\times \left(4\pi\,r_s^2\right)\times dE\times\left(\frac{a(t)}{a(t_0)}\right)^4=\frac{\pi\,g_h\,\mbh(t)^2\,n_{\rm BH}(t)}{2\,M_P^4}\,\frac{f_0^3\,df_0\,dt}{\exp(2\pi\,f/\Tbh)-1}\,,    
\end{align}
where $r_s=\mbh/(4\pi\,M_P^2)$ is the Schwarzschild radius and $g_h=0.05$~\cite{Hooper:2020otu} is the effective degrees of freedom of graviton. The ratio of scale factors takes into account the effect of redshift. Assuming no further entropy injection from the end of PBH evaporation till today, the total energy density at present
\begin{align}
& d\rho(f_0)=\frac{g_h\,\pi}{2\,M_P^4}\,\int_{\tin}^{\tev}df_0\,dt\,f_0^3\,\frac{\mbh(t)^2\,n_{\rm BH}(t)}{\exp\left[2\pi\,f_0\,\mbh(t)/(a(t)\,M_P^2)\right]-1}\,.    
\end{align}
The spectral energy density of the GW is then
\begin{align}
& \ogw(f_0)=\frac{1}{\rho_c}\,\frac{d\rho(f_0)}{d\ln f_0}=\frac{g_h\,\pi}{2\,M_P^4\,\rho_c}\,f_0^4\,\int_{\tin}^{\tev}dt\,\frac{\mbh(t)^2\,n_{\rm BH}(t)}{\exp\left[2\pi\,f_0\,\mbh(t)/(a(t)\,M_P^2)\right]-1}
\nonumber\\&\equiv
\frac{g_h\,\pi}{2\,M_P^4\,\rho_c}\,f_0^4\,\int_{\ain}^{\aev}\,\frac{da}{a^2\,H}\,\mathcal{Q}(t,f_0)\,,    
\end{align}
where $\rho_c \simeq 1.05 \times 10^{-5}\, h^2$~GeV/cm$^3$ is the critical energy density of the Universe. The PBH mass evolves as~\cite{Haque:2024eyh,Barman:2024iht}
\begin{align}
& \mbh(t)=
\begin{cases}
\Min[1-\Gamma_{\rm BH}^0(t-t_{\rm in})]^{1/3} & t < t_q\,,
\\ 
q\,\Min\l[1-\Gamma_{\rm BH}^k(t-t_q)\r]^{\frac{1}{3+2k}} & t \geq t_q\,,
\end{cases}
\end{align}
with the BH numberd density 
\begin{align}
& n_{\rm BH}=n_{\rm BH}^{\rm in}\,\left(\ain/a\right)^3\,,  
\end{align}
where $n_{\rm BH}^{\rm in}$ is given by Eq.~\eqref{eq:nbhin}.

\bibliography{Bibliography}
\bibliographystyle{JHEP}
\end{document}